\documentclass[preprint,12pt]{elsarticle}

\usepackage{amssymb}
\usepackage{graphicx}

\journal{Solid State Communications}

\begin{document}

\begin{frontmatter}


\title{The electronic structure of CeNiGe$_3$ and YNiGe$_3$ superconductors by {\it ab initio} calculations}

\author[INT]{M.J. Winiarski}
\author[INT]{M. Samsel-Czeka{\l}a}
\address[INT]{Institute of Low Temperature and Structure Research, Polish Academy of Sciences, Ok{\'o}lna 2, 50-422 Wroc{\l}aw, Poland}

\begin{abstract}
Band structures of pressure-induced CeNiGe$_3$ and exotic BCS-like YNiGe$_3$ superconductors have been calculated employing the full-potential local-orbital code. Both the local density approximation (LDA) and LDA+$U$ treatment of  the exchange-correlation energy were used. The investigations were focused on differences between electronic properties of both compounds. Our results indicate that the Ce-based system exhibits higher density of states at the Fermi level, dominated by the Ce 4f states, in contrast to its non f-electron counterpart. The Fermi surface (FS) of each compound originates from three bands and consists of both holelike and electronlike sheets. The specific FS nesting properties of only CeNiGe$_3$ enable an occurrence of spin fluctuations of a helicoidal antiferromagnetic character that may lead to unconventional pairing mechanism in this superconductor. In turn, the topology of the FS in YNiGe$_3$ reveals a possibility of multi-band superconductivity, which can explain 
the observed anomalous jump at T$_c$ in its specific heat.
\end{abstract}


\end{frontmatter}

\section{Introduction}

Among Ce-Ni-Ge intermetallics, the pressure-induced unconventional superconductivity (SC) is exhibited by e.g. Ce$_2$Ni$_3$Ge$_5$ \cite{ref1,ref2} and CeNi$_2$Ge$_2$ \cite{ref3,ref4}, however, the highest superconducting  transition temperature T$_c$ (= 0.48 K) was reported for studied in this paper CeNiGe$_3$ \cite{ref5}. It is known that the phenomenon of pressure-induced SC in such f-electron systems occurs around the quantum critical point (QCP), where their antiferromagnetic (AFM) order vanishes and the heavy-fermion (HF) state is formed. Meanwhile, in CeNiGe$_3$, an SC mechanism is more complex, since its SC is robust not only in the HF state (with strong spin fluctuations) at QCP for critical pressure of 6-7 GPa, but in a much wider pressure range (1.7 - 9.3 GPa), where it is able also to coexist with an ordered AFM state  \cite{ref6,ref7,ref8}.

The discovery of non-f-electron YNiGe$_3$ counterpart of CeNiGe$_3$ enables a comparison between electronic structures of these two isostructural orthorhombic superconductors. Furthermore, the T$_c$=0.46 K \cite{ref9} in the considered here Y-based compound is close to that in its Ce-based analogue, but the character of SC in the former system is rather BCS-like with some multi-band anomalies. Investigations of electronic properties of these two systems may be helpful in understanding differences between their SC features.
 
Magnetic properties of CeNiGe$_3$ \cite{ref10,ref11} have been investigated with focus on their similarities to those in Ce$_3$Ni$_2$Ge$_7$ by both experimental \cite{ref12} and theoretical methods \cite{ref13}. The competition between collinear and helicoidal, with the propagation vector {\bf k}=(0, 0.409, 1/2), AFM ground states was reported for CeNiGe$_3$ \cite{ref11}. In turn, the considered isostructural Y-based system has also been used as a non-magnetic counterpart to CeNiGe$_3$ \cite{ref14,ref15}. Despite the fact that the substitution of Ce with Y atoms results in clear reduction of the unit cell (u.c.) volume, acting as high hydrostatic pressure, the most apparent discrepancies between CeNiGe$_3$ and YNiGe$_3$ can originate from their electronic structure, connected with the presence of the Ce 4f electrons only in the former compound.

In YNiGe$_3$ an anomalous jump in specific heat, $\Delta C/ \gamma T$ = 0.5 at T$_c$ = 0.46 K, with respect to the standard BCS value (1.43), clearly revealed an exotic character of  SC in this germanide. Also other Ni-Ge rare-earth superconductor, La$_3$Ni$_4$Ge$_4$, exhibits rather a weak-coupling behavior, but its $\Delta C/ \gamma T$ = 0.95 at T$_c$= 0.7 K \cite{ref16} is significantly higher than the jump in YNiGe$_3$. It is suspected that only very high anisotropy of the superconducting gap may lead to such a low value of the specific heat jump at a raise of SC as in the case of two-gap SC in MgB$_2$ for which also a much reduced jump $\Delta C/ \gamma T$ = 0.82 at T$_c$ has been reported \cite{ref17}. 

In this work, band structures of both CeNiGe$_3$ and YNiGe$_3$ ternaries are investigated from first principles within the DFT (density functional theory) framework. The main aim of this study is the comparison between the topology of their Fermi surfaces and values of the density of states (DOS) at the Fermi level (E$_F$). The analysis of similarities of the studied here systems to other $R$-Ni-Ge (where $R$ is a rare-earth atom) intermetallics is also carried out.

\section{Computational details}

Electronic structure calculations for CeNiGe$_3$ and YNiGe$_3$, crystallizing in the orthorhombic SmNiGe$_3$-type structure ({\it Cmmm}, space group no. 65), have been performed with the full-potential local-orbital (FPLO-9) method \cite{ref18}. The Perdew-Wang form of the local density approximation (LDA) of the exchange-correlation functional \cite{ref19} was employed in the scalar relativistic mode. Usually strongly-correlated Ce 4f electrons were additionally treated within the LDA+$U$ approach, assuming a commonly accepted value of the effective Coulomb repulsion potential $U$ = 6 eV that has been employed also in other Ce-Ni-Ge \cite{ref20} and Ce-based intermetallics \cite{ref21}. For CeNiGe$_3$, the experimental neutron powder diffraction data of lattice parameters {\it of the u.c. volume, $V_0$, at ambient pressure}:  $a$ = 2.18083, $b$ = 0.41351, $c$ =0.41684 nm and atomic positions in the u.c.: Ce in (4h): (0.1681, 0, 1/2); Ni in (4g): (0.39128, 0, 0); Ge(1) in (4g): (0.28331, 0, 0); Ge(2) in (4g):
 (0.0564, 0, 0); Ge(3) in (4h): (0.4439, 0, 1/2) have been taken from Ref. \cite{ref10}. 
{\it In addition, two sets of proportionally compressed lattice parameters were assumed to simulate possible changes of electronic structure, and particularly the FS, under hydrostatic pressure. The assumed two sets of parameters corresponded to reduced u.c. volumes: 99\%$V_0$ and 98\%$V_0$.}
In turn, for YNiGe$_3$, the measured x-ray diffraction values of lattice parameters: $a$ = 2.1529, $b$ = 0.4060, and $c$ = 0.4063 nm \cite{ref9} were utilized. Since the lack of experimental atomic positions for YNiGe$_3$, the above ones for the Ce-based compound were assumed. Here the u.c. is equivalent to double formula units (f.u.). Valence-basis sets were automatically selected by the internal procedure of FPLO-9. Total energy values of considered systems were converged with accuracy to ~1 meV for the 16x16x16 {\bf k}-point mesh of the Brillouin zone (BZ), corresponding to 657 {\bf k}-points in the irreducible part of the BZ.

\section{Results and discussion}

\begin{figure}
\includegraphics[scale=1.0]{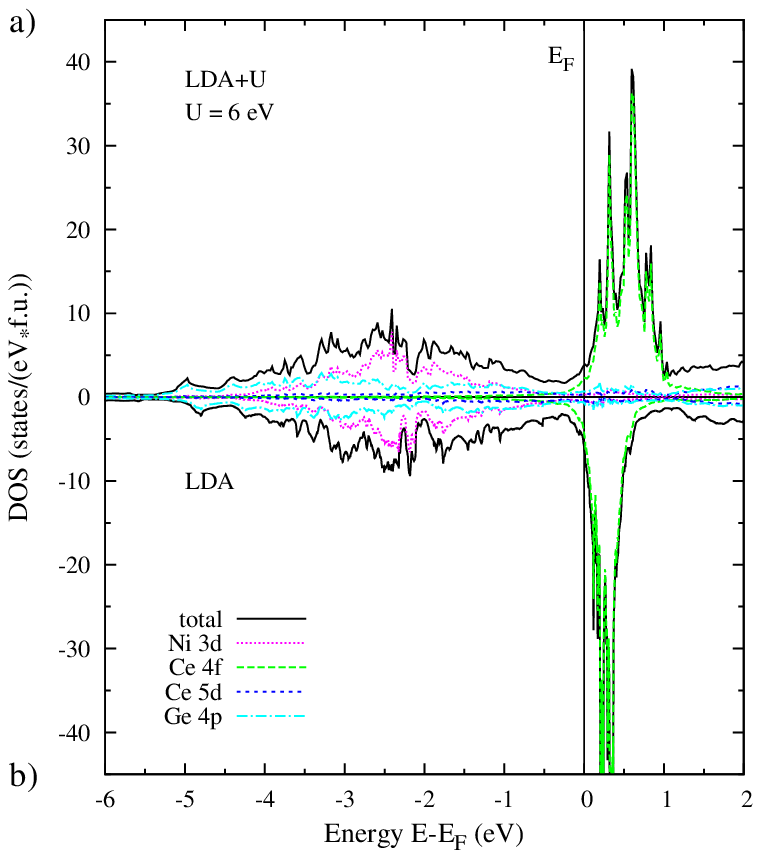}
\caption{Total
 and partial (per electron orbitals) DOS plots for CeNiGe$_3$, calculated employing: a) LDA+$U$ with $U$ = 6 eV and b) standard LDA approaches.}
\label{Fig1}
\end{figure}

In spite of the character of the Ce 4f electrons in CeNiGe$_3$ that has been considered rather as itinerant in Ref. \cite{ref13}, the inclusion of the repulsive Coulomb potential $U$ may yield a more adequate description of a normal state in this system. For this germanide superconductor, our calculated DOS plots, obtained by both LDA and LDA+$U$ approaches are presented in Fig. \ref{Fig1}. As seen in this figure, the overall shape of DOS contributions, originating from Ni and Ge atoms, appears to be almost unchanged by the inclusion of $U$ = 6 eV for the Ce 4f states, being also similar to the LDA results for Ce$_3$Ni$_2$Ge$_7$, reported earlier \cite{ref13}. The total DOS at the Fermi level, N($E_F$), is dominated by the Ce 4f electrons. However, the LDA+$U$ result of N(E$_F$) reveals a remarkable decrease from 6.0 (for LDA) to 3.75 states/eV/f.u. It is an effect of the shift of the Ce 4f peak to higher energies above E$_F$ after including the Coulomb potential $U$= 6 eV. Moreover, the shifted peak is 
about two times broader than that obtained with the LDA approach. 
{\it These values remain unchanged with hydrostatic pressure considered here.}
Furthermore, earlier specific heat measurements revealed the Sommerfeld coefficient $\gamma_0$ = 34 mJ Ce/mol/K$^2$ for CeNiGe$_3$ \cite{ref11}, being almost three times lower than $\gamma_0$ (= 90 mJ Ce/mol/K$^2$) for Ce$_2$Ni$_3$Ge$_5$
 \cite{ref22}. Meanwhile, for CeNiGe$_3$ the above N(E$_F$) values are only slightly lower than the corresponding LDA or LDA+$U$ results for Ce$_2$Ni$_3$Ge$_5$, equal to 6.8 or 5.0 states/eV/f.u., respectively \cite{ref23}. Thus, the enhancement of the experimental $\gamma_0$, compared with the theory, is stronger in the latter superconductor.  

\begin{figure}
\includegraphics[scale=1.0]{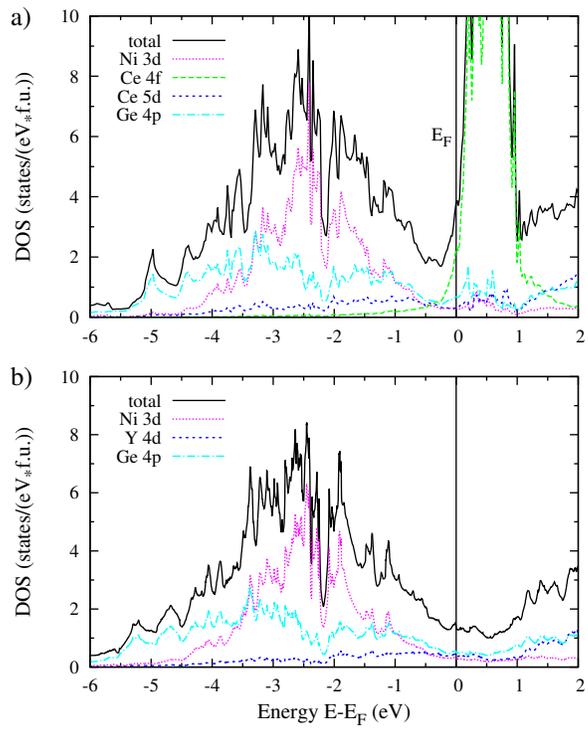}
\caption{Calculated total and partial DOS plots for: a) CeNiGe$_3$ (LDA+$U$) and b) YNiGe$_3$ (LDA).}
\label{Fig2}
\end{figure}

The lack of the 4f electrons in YNiGe$_3$ results in much lower value of N(E$_F$) = 1.38 states/eV/f.u. than that in the above germanides, as illustrated in Fig. \ref{Fig2}. Moreover, the total DOS at E$_F$ in this Y-based compound is formed by equal contributions, coming from the Ni 3d, Y 4d and Ge 4p electrons, analogously to those in other orthorhombic Y-Ni intermetallics, e.g. Y$_2$Ni$_3$Si$_5$ with N(E$_F$) = 2.5 states/eV/f.u. \cite{ref24}. It is worth to notice that in CeNiGe$_3$, N(E$_F$) without the contributions of the Ce 4f electrons is almost equal to that in YNiGe$_3$ (see Fig. \ref{Fig2}). Nevertheless, the similar values of T$_c$'s ($\approx$ 0.5 K) in both compounds seem to be rather coincidental, taking also into account the HF character of SC only in CeNiGe$_3$, exhibited under pressure.

\begin{figure}
\includegraphics[scale=0.85]{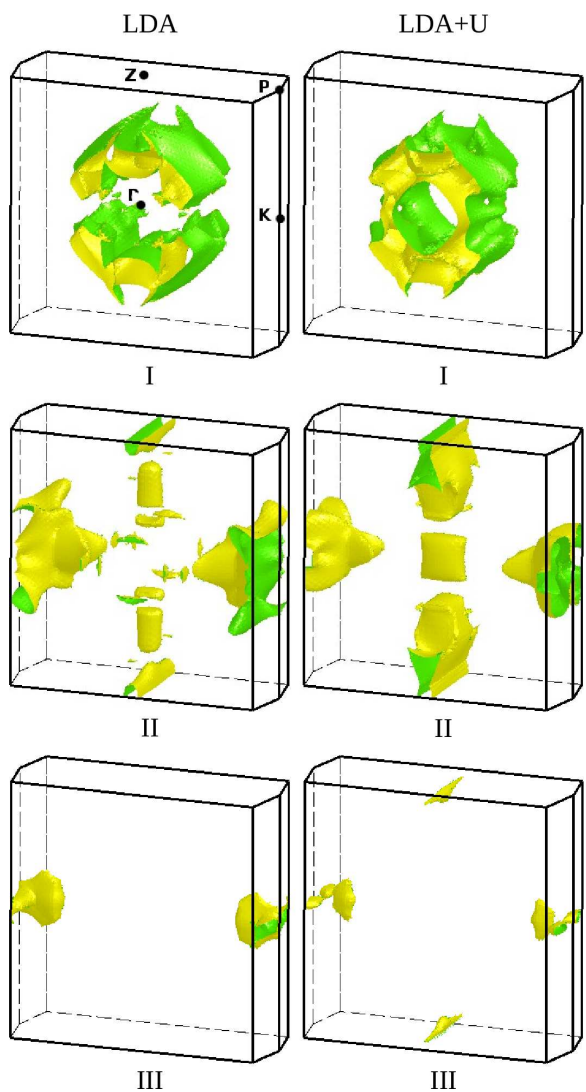}
\caption{The Fermi surface sheets of CeNiGe$_3$, originating from three conduction bands (denoted as I-III), computed by LDA and LDA+$U$ approaches (left and right panels, respectively) and drawn in the orthorhombic BZ boundaries.}
\label{Fig3}
\end{figure}

The FS of CeNiGe$_3$ consists of three sheets, originating from three conduction bands denoted as I-III. The largest holelike sheet I, centered around the $\Gamma$ point, exhibits quasi-2D character along the elongated $a$ axis, while the two remaining FS sheets are built from smaller 3D electronlike closed pockets, as depicted in Fig. \ref{Fig3}. The differences between the LDA and LDA+$U$ results of FS are rather subtle, however, the latter exhibit more cylindrical character of FS sheet I and a lack of artefactual tiny pieces in FS sheet II.
{\it Furthermore, the FS sheets after simulating hydrostatic pressure (not shown) would be the same in the scale of Fig. \ref{Fig3}.}

\begin{figure}
\includegraphics[scale=1.0]{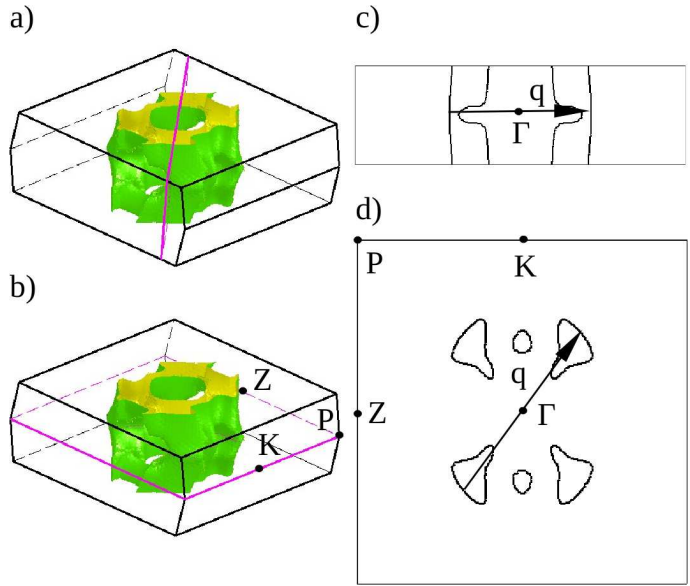}
\caption{Two sections, a) - c) and b) - d), through Fermi-surface sheet I of CeNiGe$_3$ (LDA+$U$) with the imperfect nesting vector $\mathbf q=(0, 0.409, 1/2)$, marked by arrows.}
\label{Fig4}
\end{figure}

\begin{figure}
\includegraphics[scale=0.8]{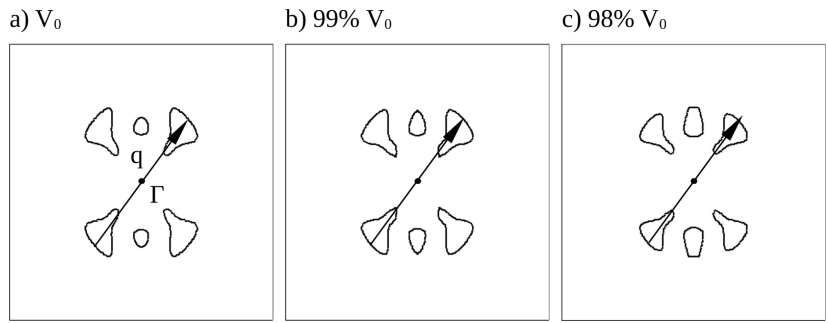}
\caption{The same as in Fig. 4 d) but for the u.c. volume: a) at ambient pressure, $V_0$, and under pressure, corresponding to: b) 99\%$V_0$ and c) 98\%$V_0$.}
\label{Fig5}
\end{figure}

More detailed investigations of the FS in CeNiGe$_3$ have revealed the unique nesting properties, displayed in Fig. \ref{Fig4}. The imperfect nesting vector {\bf q} in FS sheet I is in very good accord with the wave vector {\bf k}=(0, 0.409, 1/2) of the helicoidal AFM order, observed experimentally \cite{ref11}. 
{\it It is worth underlining that both the dimensions and topology of the whole FS, and especially the nested FS piece, remain almost the same for two sampling values of hydrostatic pressure, as illustrated in Fig. \ref{Fig5}. Only at pressure corresponding to the volume 98\%$V_0$ the nesting vector becomes even less perfect in its length. Therefore, our findings do not indicate an FS reconstruction scenario (changes of the FS volume connected with f-electron localization-delocalization process) under the probed hydrostatic pressure.}
Interestingly, a nesting vector, also being consistent with the wave vector of an AFM structure, was reported for other HF superconductor under pressure, Ce$_2$Ni$_3$Ge$_5$ \cite{ref23} among Ce-Ni-Ge systems. Hence, in these Ce-Ni-Ge intermetallics, {\it slight modifications of} the imperfect nesting properties {\it by pressure} may influence the strength of spin fluctuations of an AFM-type that could be responsible for the unconventional SC pairing mechanism in these systems under pressure. In CeNiGe$_3$, in which two types of AFM interactions compete with each other \cite{ref11}, spin fluctuations of the helicoidal-type can be present {\it also in a broader} lower-pressure region, where SC coexists with the AFM long-range ordering of possibly a 
collinear-type. However, this issue requires further experimental verifications.

\begin{figure}
\includegraphics[scale=0.8]{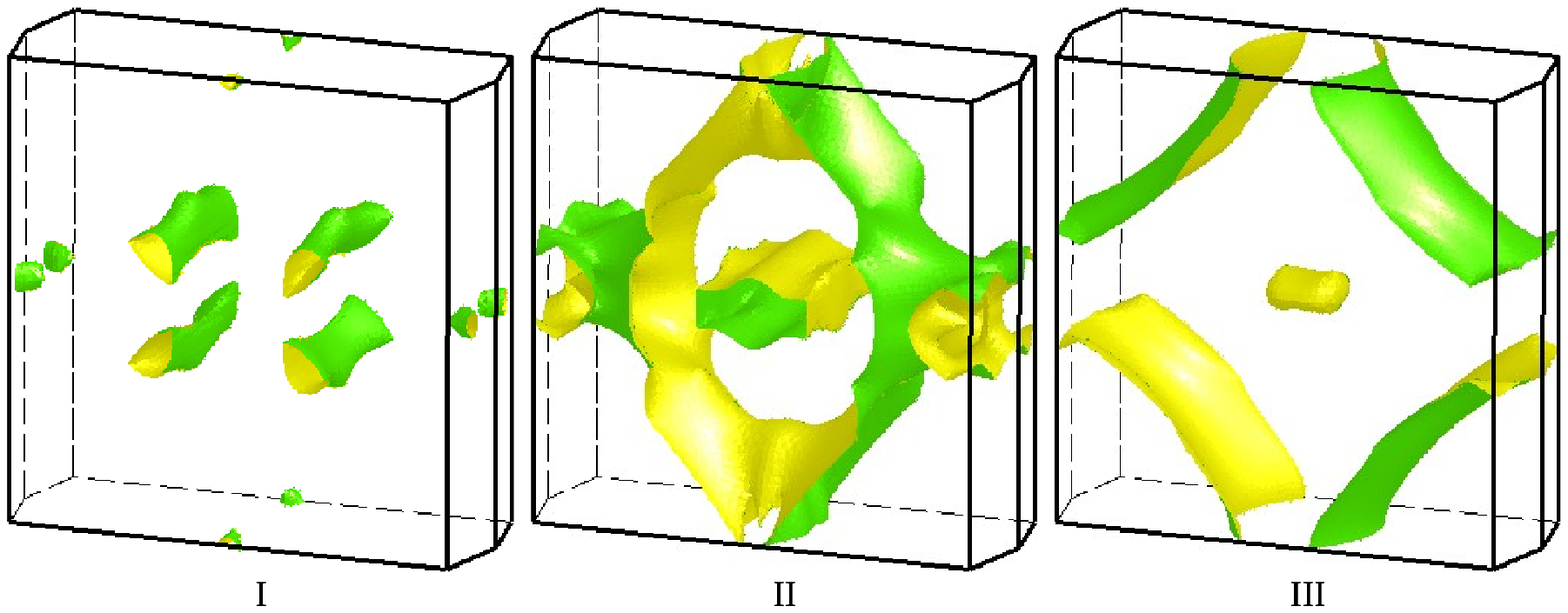}
\caption{The calculated (LDA) Fermi surface sheets of YNiGe$_3$, coming from three bands (labeled as I-III) in the orthorhombic BZ boundaries.}
\label{Fig6}
\end{figure}

In turn, the FS of YNiGe$_3$, existing also in three bands (labeled as I-III), is completely different from that in CeNiGe$_3$, as visualized in Fig. \ref{Fig6}. FS sheet I contains small holelike quasi-2D pipes (along the $a$ axis), located around the $\Gamma$ point. Also FS sheet III consists of quasi-2D pieces but having larger cylindrical shapes, placed in the corners of the BZ. In turn, FS sheet II is more 3D and complex than the above ones. It is worth to note that in other non f-electron orthorhombic Ni-Ge system, La$_3$Ni$_4$Ge$_4$  \cite{ref25}, the numerous FS sheets are very similar to those in YNiGe$_3$. The specific heat jump $\Delta C/ \gamma T$ = 0.95 at T$_c$= 0.7 K \cite{ref16} in the 344-type system may be connected with an anisotropy of the SC gaps opened on these FS sheets of different dimensionality. However, the anomalous value of $\Delta C/ \gamma T$ = 0.5 at T$_c$ in YNiGe$_3$ is significantly lower, especially in comparison with other multi-band superconductors, e.g. MgB$_2$ ($\Delta 
C/ \gamma T$ = 0.82 at T$_c$ \cite{ref17}). These findings suggest a multi-band character of SC in YNiGe$_3$ with strongly anisotropic SC gaps. One can consider two SC gaps, opened on FS sheets II and III, differing in their dimensionality. However, further experimental studies on  the issue of SC gaps in YNiGe$_3$ are necessary to confirm these suggestions.

\section{Conclusions}

The electronic structures of CeNiGe$_3$ and YNiGe$_3$ superconductors have been investigated from first principles. The density of states at the Fermi level in the Ce-based compound is dominated by the Ce 4f electrons while its Fermi surface exhibits imperfect nesting properties, {\it conserved under hydrostatic pressure, being} consistent with the wave vector of the helicoidal AFM structure. The possibility of tuning spin fluctuations by this nesting {\it under pressure} may be crucial for unconventional superconductivity in this compound. Meanwhile, the electronic structure of its isostructural counterpart, YNiGe$_3$, is considerably different. The multi-band character of rather BCS-like superconductivity in YNiGe$_3$ is possible due to its Fermi surface topology. These findings may explain an anomalous value of the specific-heat jump (= 0.5) at T$_c$ in YNiGe$_3$, reported earlier \cite{ref9}.

\section*{Acknowledgments}
This work has been supported by the National Center for Science in Poland (Grant No. N N202 239540). The calculations were performed in Wroc{\l a}w Center for Networking and Supercomputing (Project No. 158).

\begin{thebibliography}{99}

\bibitem{ref1} 
M. Nakashima, H. Kohara, A. Thamizhavel, T.D. Matsuda, Y. Haga, M. Hedo , Y. Uwatoko, R. Settai, Y. \={O}nuki, J. Phys.: Condens. Matter 17 (2005) 4539.

\bibitem{ref2} 
M. Nakashima, H. Kohara, A. Thamizhavel, T.D. Matsuda, Y. Haga, M. Hedo, Y. Uwatoko, R. Settai, Y. \={O}nuki, Physica B 378-380 (2006) 402. 

\bibitem{ref3} 
F.M. Grosche, S.J.S. Lister, F.V Carter, S.S Saxena, R.K.W. Haselwimmer, N.D Mathur, S.R. Julian, G.G. Lonzarich, Physica B 239 (1997) 62.

\bibitem{ref4} 
F.M. Grosche, P. Agarwal, S.R. Julian, N.J. Wilson, R.K.W. Haselwimmer, S.J.S. Lister, N.D. Mathur, F.V. Carter, S.S Saxena, G.G. Lonzarich, J. Phys.: Condens. Matter 12 (2000) L533.

\bibitem{ref5} 
M. Nakashima, K. Tabata, A. Thamizhavel, T.C. Kobayashi, M. Hedo, Y. Uwatoko, K. Shimizu, R. Settai, Y. \={O}nuki, J. Phys.: Cond. Matt. 16 (2004) L255. 

\bibitem{ref6} 
M. Nakashima, K. Tabata, A. Thamizhavel, T.C. Kobayashi, M. Hedo, Y. Uwatoko, K. Shimizu, R. Settai, Y. \={O}nuki, Physica B 359-361 (2005) 266.

\bibitem{ref7} 
H. Kotegawa, T. Miyoshi, K. Takeda, S. Fukushima, H. Hidaka, K. Tabata, T.C. Kobayashi, M. Nakashima, A. Thamizhavel, R. Settai, Y. \={O}nuki, Physica B 378-380 (2006) 419.

\bibitem{ref8} 
H. Kotegawa, K. Takeda, T. Miyoshi, S. Fukushima, H. Hidaka, T.C. Kobayashi, T. Akazawa, Y. Ohishi, M. Nakashima, A. Thamizhavel, R. Settai, Y. \={O}nuki, J. Phys. Soc. Jpn. 75 (2006) 044713.

\bibitem{ref9} 
A.P. Pikul, D. Gnida, Solid State Commun. 151 (2011) 778.

\bibitem{ref10} 
L. Durivault, F. Bour{\'e}e , B. Chevalier, G. Andr{\'e}, F. Weill, J. Etourneau, Appl. Phys. A 74 (2002) S677.

\bibitem{ref11} 
L. Durivault, F. Bour{\'e}e, B. Chevalier, G. Andr{\'e}, F. Weill, J. Etourneau, P. Martinez-Samper, J.G. Rodrigo, H. Suderow, S. Vieira, J. Phys.: Condens. Matter 15 (2003) 77.

\bibitem{ref12} 
L. Durivault, F. Bouree, B. Chevalier, G. Andre, J. Etourneau, O. Isnard, J. Magn. Magn. Mater. 232 (2001) 139.

\bibitem{ref13} 
S.F. Matar, B. Chevalier, O. Isnard, J. Etourneau, J. Mater. Chem. 13 (2003) 916.

\bibitem{ref14} 
A. P. Pikul, D. Kaczorowski, T. Plackowski, A. Czopnik, H. Michor, E. Bauer, G. Hilscher, P. Rogl, Yu. Grin, Phys. Rev. B 67 (2003) 224417.

\bibitem{ref15} 
A.P. Pikul, D. Kaczorowski, P. Rogl, Yu. Grin, Phys. Stat. Sol. B 236 (2003) 364.

\bibitem{ref16} 
S. Kasahara, H. Fujii, H. Takeya, T. Mochiku, A.D. Thakur, K. Hirata, J. Phys.: Condens. Matter 20 (2008) 385204.

\bibitem{ref17} 
Y. Wang, T. Plackowski, A. Junod, Physica C 355 (2001) 179.

\bibitem{ref18}
FPLO9.00-34, improved version of the FPLO code by K. Koepernik, H. Eschrig, Phys. Rev. B 59 (1999) 1743; www.FPLO.de.

\bibitem{ref19}
J.P. Perdew, Y. Wang, Phys. Rev. B 45 (1992) 13244.

\bibitem{ref20} 
C. de la Fuente, A. del Moral, D.T. Adroja, A. Fraile, J.I. Arnaudas, J. Magn. Magn. Mater. 322 (2010) 1059.

\bibitem{ref21} 
J. Goraus, A. {\'S}lebarski, M. Fija{\l}kowski, Phys. Stat. Solid. B 248 (2011) 2857.

\bibitem{ref22} 
Z. Hossain, S. Hamashima, K. Umeo, T. Takabatake, C. Geibel, F. Steglich, Phys. Rev. B 62 (2000) 8950.

\bibitem{ref23} 
M.J. Winiarski, M. Samsel-Czeka{\l}a, J. Alloys Compd. 560 (2013) 123.

\bibitem{ref24} 
M. Samsel-Czeka\l a,  M.J. Winiarski, Intermetallics 20 (2012) 63.

\bibitem{ref25} 
M.J. Winiarski, M. Samsel-Czeka{\l}a, J. Alloys Compd. 546 (2013) 124. 

\end{thebibliography}
\end{document}